\documentclass[onecolumn,preprintnumbers,pre]{revtex4}
\usepackage{graphicx, subfigure}
\usepackage{amssymb, amsmath,amssymb,amsfonts}
\usepackage{amsthm,mathrsfs,amsopn}
\usepackage{dcolumn}
\usepackage{bm}
\usepackage{color}
\usepackage[utf8]{inputenc}

\theoremstyle{plain} 

\begin{document}

\title{How to fairly share a watermelon}

\author{ Timoteo Carletti$^1$}
\affiliation{$^1$naXys, Namur Institute for Complex Systems, University of Namur, rempart de la vierge 8, B5000 Namur, Belgium}

\author{Duccio Fanelli$^2$}
\affiliation{$^2$ Universit\`{a} degli Studi di Firenze, Dipartimento di Fisica e Astronomia,
CSDC and INFN, via G. Sansone 1, 50019 Sesto Fiorentino, Italy}

\author{Alessio Guarino$^3$}
\affiliation{$^3$ Universit\'e de La R\'eunion – Laboratoire Icare EA 7389, France}

\begin{abstract}
Geometry, calculus and in particular integrals, are too often seen by young students as technical tools with no link to the reality. This fact generates into the students a loss of interest with a consequent removal of motivation in the study of such topics and more widely in pursuing scientific curricula. With this note we put to the fore a simple example of practical interest where the above concepts prove central; our aim is thus to motivate students and to {reverse} the dropout trend by proposing an introduction to the theory starting from practical applications. More precisely, we will show how using a mixture of geometry, calculus and integrals one can easily share a watermelon into regular slices with equal volume.
\end{abstract}

\maketitle

\section{Introduction to the main question}
\label{sec:intro}

What better than a fresh watermelon to fight the heat wave in these sunny and hot summer days? In Italy it is very common to buy watermelons weighting as much as $15\, \mathrm{Kg}$ for few euros. Sharing the watermelon is hence an opportunity for a party with many friends. A recurring question is then: {\em how can one slice the watermelon into equal volume parts to have a fair sharing among friends?} This question is particularly interesting if the slices are not cut along the longitudinal direction, namely the longer one. Cuts along the transversal direction are in fact often performed to return slices which can be more straightforwardly manipulated. In the following we will show that symmetry helps: for the first cuts up to a certain number of friends, one can indeed proceed by dividing each portion into two identical parts. To go further, when the number of guests is large, we propose a rule of the thumb which follows a simple mathematical analysis: this will be referred as the  ``$2/3$ rule''. {Our ultimate goal is to put forward a simple application of pedagogical interest, which can motivate students to familiarise with integral calculus and other technical tools routinely employed in physics and mathematics~\cite{WL1984,almalaurea,nytimes,istat2011,Boilevin2013,TFHB2001,MLS2020}.}

\begin{figure}[ht]
\centering
\includegraphics[scale=0.3]{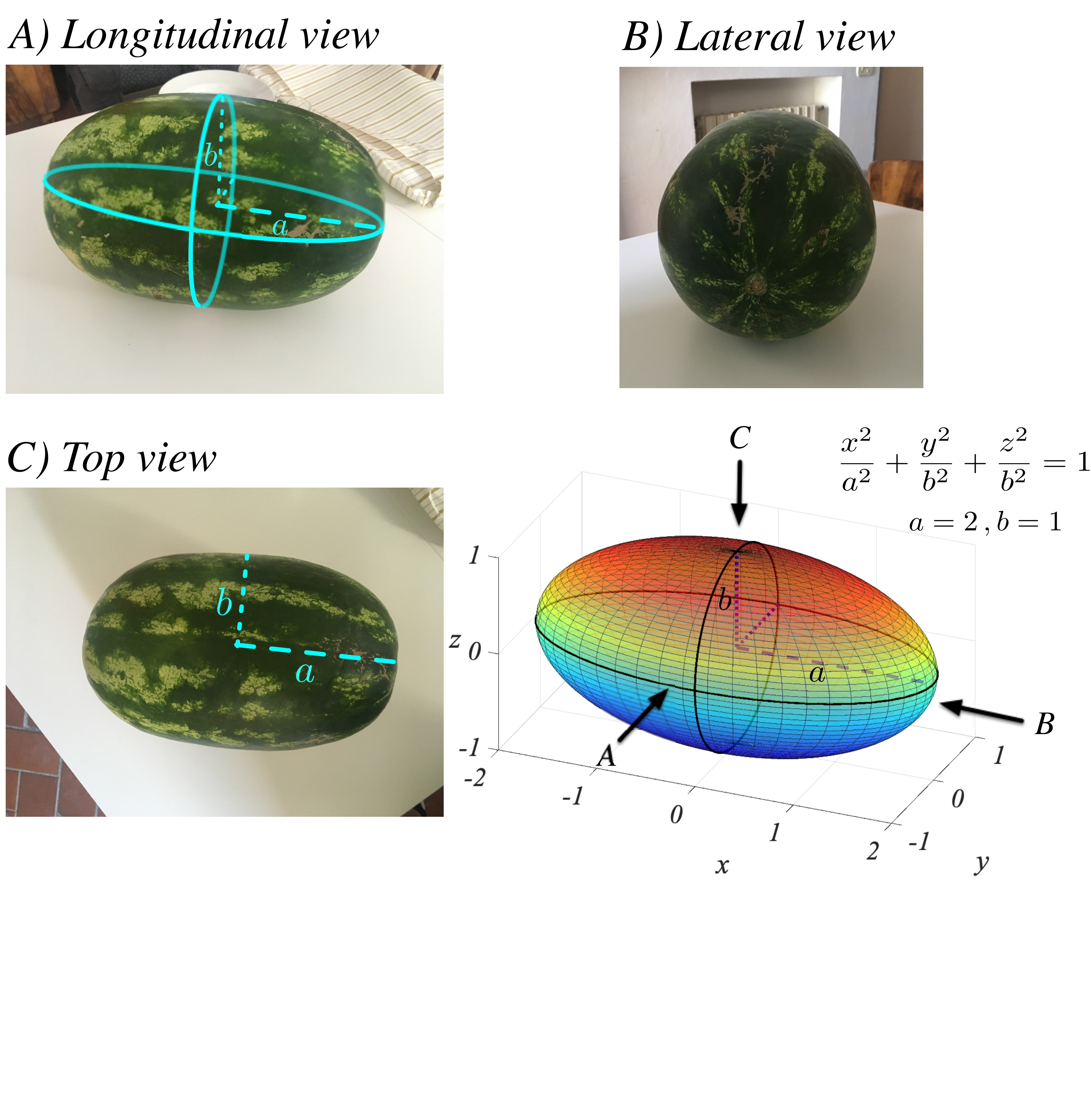}
\vspace{-2.5cm}
\caption{\textbf{Watermelon}. Different views of a watermelon (panels A, B and C). Bottom right: a (spheroid) ellipsoid with equal median and semi-minor axes, $b$, and semi-major axis, $a>b$ whose cartesian equation is given by $x^2/a^2+y^2/b^2+z^2/b^2=1$. The geometrical shape depicted in the bottom right panel corresponds to the choice $a=2$ and $b=1$ in arbitrary units. {To help visualising the geometrical setting, we show in panels A, C, as well as in the bottom right panel, the semi-major axis $a$ and the semi-minor axis $b$.}}
\label{fig:cocomero1}
\end{figure}

\begin{figure}[ht]
\centering
\includegraphics[scale=0.3]{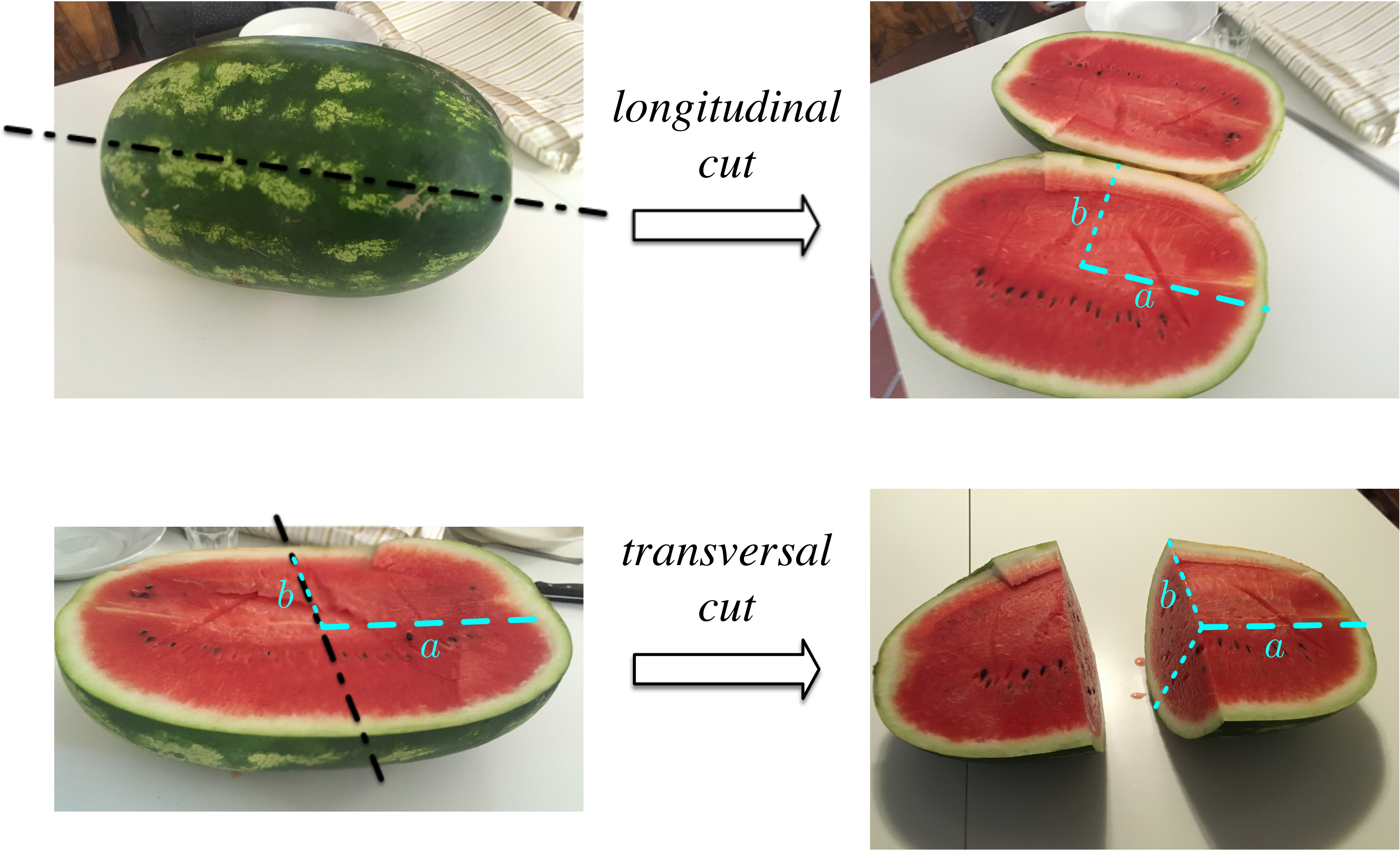}
\caption{\textbf{First few cuts}. Cutting along the symmetry axes of the watermelon can easily {return} $4$ equal volume pieces {that will constitute the starting point for successive cuts}. {The semi-major axis $a$ and the semi-minor axis $b$ are displayed in Figure.}}
\label{fig:cocomero2}
\end{figure}

The watermelon can be accurately represented as an ellipsoid (see Fig.~\ref{fig:cocomero1}); more precisely, it can be assimilated to a spheroid{, namely an ellipsoid} with two equal semi axes, in our case the median and the minor ones, with length $b$. The major axis measures $a$, with $a>b$ {(See Figs.~\ref{fig:cocomero1} and~\ref{fig:cocomero2})}. The total volume of the watermelon is thus $V_{tot}=4\pi b^2 a /3$~\cite{DFN1984}. Our goal is to equally share it among the friends. The greengrocer straightforwardly carries out the first few cuts following the symmetries of the watermelon / spheroid, i.e. cutting along the semi-major axis (longitudinal cut) and then the semi-minor axis (transversal cut) (see Fig.~\ref{fig:cocomero2}). 

\begin{figure}[ht]
\centering
\includegraphics[scale=0.3]{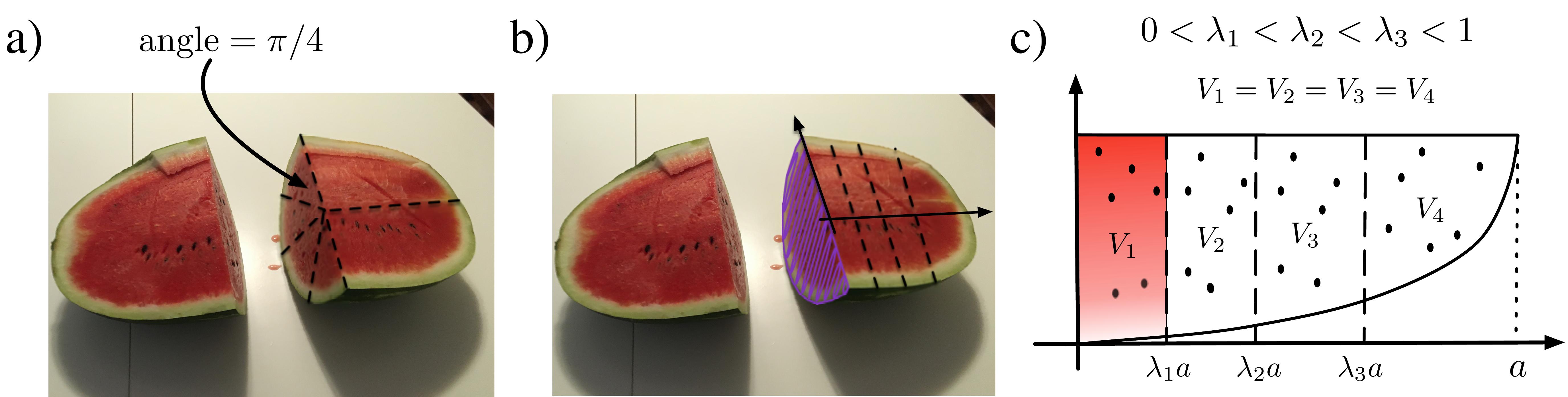}
\caption{\textbf{Where to cut next?} Panel a): cutting along the longitudinal direction by dividing the angle $\pi$ into $n$ equal parts ($n=4$ in the panel) can be hard. Panel b): cutting along the transversal direction, i.e. parallel to the central face (hatched in the figure), requires identifying the cuts positions, $0<\lambda_1 a<\lambda_2a<\dots <\lambda_{n-1} a<a$ (again $n=4$ in the panel). Panel c): the slices obtained by cutting at the positions given by the procedure presented in panel b) should all have the same volume.}
\label{fig:cocomero3}
\end{figure}

Arrived at home we {want to} make more {(equal volume)} slices from each fourth of the watermelon to serve all the friends. How to proceed? Cutting each piece along the longitudinal directions is not very easy with more than $4$ friends (see panel a) of Fig.~\ref{fig:cocomero3}). As a viable alternative{, that we shall hereafter consider,} one {can cut} the watermelon along the transversal direction{. This amounts to performing cuts parallel to the central face (hatched in panel b) of Fig.~\ref{fig:cocomero3})}. But where to cut? Equally spaced cuts do not provide the solution of the problem because the slices closer to the central face (see again panel b) Fig.~\ref{fig:cocomero3}) {bear} a larger volume. Assume we want to equally share one fourth of the watermelon between $n$ friends: to obtain $n$ slices we need to perform $n-1$ cuts. {In practice,} we have to determine the positions $0<\lambda_1 a<\lambda_2a<\dots <\lambda_{n-1} a<a$, along the semi-major axis where to slice, in such a way that each {obtained} part {shares} the same volume{. This latter should hence equal} $1/n$ of the volume of one fourth of the watermelon, $V=\pi b^2 a /3=V_{tot}/4$ (see Fig.~\ref{fig:cocomero3} panels b) and c) in the case $n=4$): 
\begin{quotation}
\textbf{Problem}: find $\lambda_i \in (0,1)$, $i=1,\dots, n-1$, such that $0<\lambda_1 <\lambda_2<\dots <\lambda_{n-1}<1$ and $V_i=V/n$, where $V_i$ is the volume of the slice {obtained by respectively cutting at positions $\lambda_{i}a$ and $\lambda_{i-1}a$}. \label{page:Vi}
\end{quotation}

\section{Computations needed to solve the problem.}
\label{sec:comput}

\begin{figure}[ht]
\centering
\includegraphics[scale=0.4]{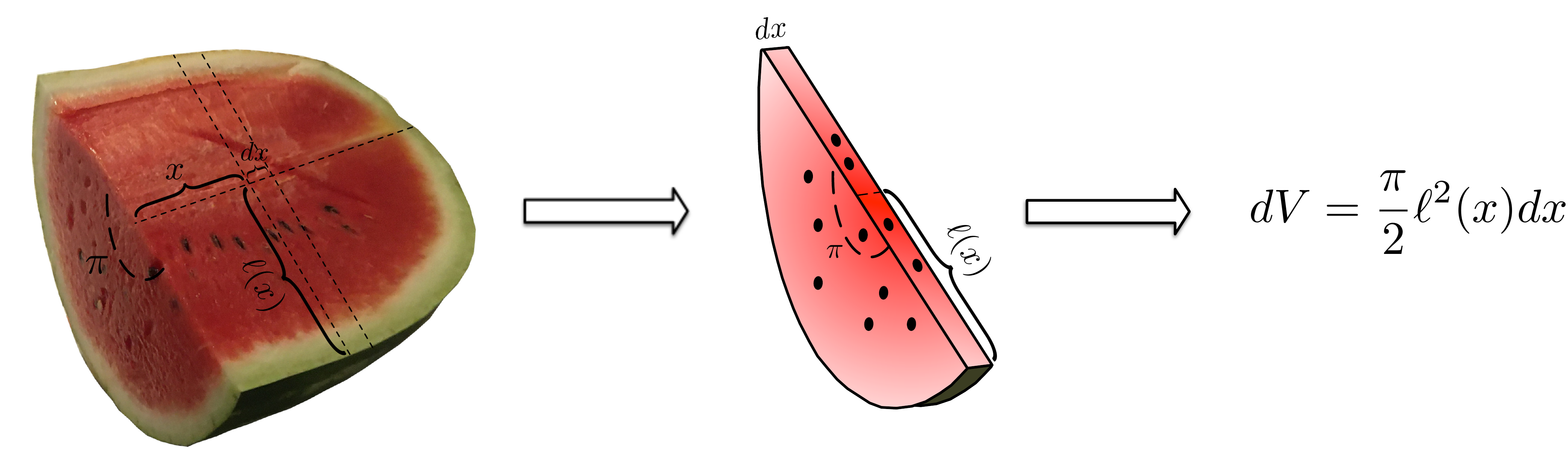}
\caption{\textbf{Geometry for the computation of the volume of the slice}. We ideally slice the large piece of watermelon into arbitrarily tiny slices of width $dx$ {(on the left panel)}. The latter can be considered as a very tiny sector of disk, with opening {$\pi$} and radius $\ell(x)$, the latter depending on the position $x$ of the slice {(on the middle panel)}. Such an arbitrary thin slice has a volume {$dV=\pi/2 \times \ell^2(x) dx$} {(on the right panel)}. Indeed it can be considered as an infinitesimally tick cylinder (with hight $dx$), whose basis is a sector of disk {with central angle $\pi$}.}
\label{fig:fetta}
\end{figure}

Let us consider $0<\lambda<1$ and compute the volume $V(\lambda a)$ of the portion of spheroid contained between the central face and the cut at distance $\lambda a$ from this face. Consider a tiny slice{, shaped as a circular sector with central (opening) angle of $\pi$, i.e. a half-disk,} at {a generic distance} $x$ {measured from the central face,} with thickness $dx$ (see Fig.~\ref{fig:fetta}). {Then,} its volume is given by $dV=\pi/2 \times \ell^2(x)dx$, where $\ell(x)=b\sqrt{1-x^2/a^2}$ is the radius of the sector of the disk. The slice is indeed a portion of a cylinder of height $dx$ and whose base is a sector of a disk with opening $\pi$ and radius $\ell(x)$; the radius depends on the position of the slice, it equals $b$ (the semi-minor and medium axis) for $x=0$ and it vanishes at $x=a$ (see panel b) of Fig.~\ref{fig:cocomero3}). In conclusion we get for the volume $V(\lambda a)$
\begin{equation}
\label{eq:Vlambda}
V(\lambda a)=\frac{\pi}{2}\int_0^{\lambda a} \ell^2(x)dx=\frac{\pi}{2}b^2\int_0^{\lambda a} \left(1-\frac{x^2}{a^2}\right)dx=\frac{\pi}{2} b^2 a \lambda \left(1-\frac{\lambda^2}{3}\right)\, .
\end{equation}

Given the integer $n\geq 2$ and $\lambda_i \in(0,1)$ such that $0<\lambda_1<\lambda_2<\dots <\lambda_{n-1} <1$, then we get for the volumes $V_i$ {(defined on page~\pageref{page:Vi})}
\begin{equation}
\label{eq:root0}
V_1=V(\lambda_1 a)\, , V_{i+1}=V(\lambda_{i+1} a)-V(\lambda_i a) \quad \forall i=1,\dots, n-2 \text{ and } V_{n}=V(a)-V(\lambda_{n-1} a) \, .
\end{equation}
Our problem is thus equivalent to solve the equations
\begin{eqnarray}
\label{eq:root}
\frac{1}{n}&=&\frac{V_1}{V}=\frac{3}{2}\lambda_1\left(1-\frac{\lambda_1^2}{3}\right)\notag\, , \\ 
\frac{1}{n}&=&\frac{V_{i}}{V}=\frac{3}{2}\lambda_{i}\left(1-\frac{\lambda_{i}^2}{3}\right)-\frac{3}{2}\lambda_{i-1}\left(1-\frac{\lambda_{i-1}^2}{3}\right) \quad \forall i=2,\dots, n-1\text{ and }\notag\\
\frac{1}{n}&=&\frac{V_{n}}{V}=\frac{V(a)}{V}-\frac{3}{2}\lambda_{n-1}\left(1-\frac{\lambda_{n-1}^2}{3}\right)\, .
\end{eqnarray}
Stated differently, let $p(\lambda)=\frac{3}{2}\lambda\left(1-\frac{\lambda^2}{3}\right)$, the first equation of~\eqref{eq:root} is equivalent to looking for a root of $p(\lambda_1)=1/n$. Then given $\lambda_1$ one can plug this value in the next equation for $i=2$ and solve for $\lambda_2$ the equation $p(\lambda_2)=1/n+p(\lambda_1)= 2/n$, where we used that fact that $p(\lambda_1)=1/n$. We can iterate the above steps and get:
\begin{equation*}
\forall i=2,\dots, n-1 \quad \frac{i}{n}=p(\lambda_{i})\, .
\end{equation*}
The last equation involving $V_n$ results into the identity $p(1)=1$, i.e. the volume of the whole piece is $V$.

\begin{figure}[ht]
\centering
\includegraphics[scale=0.4]{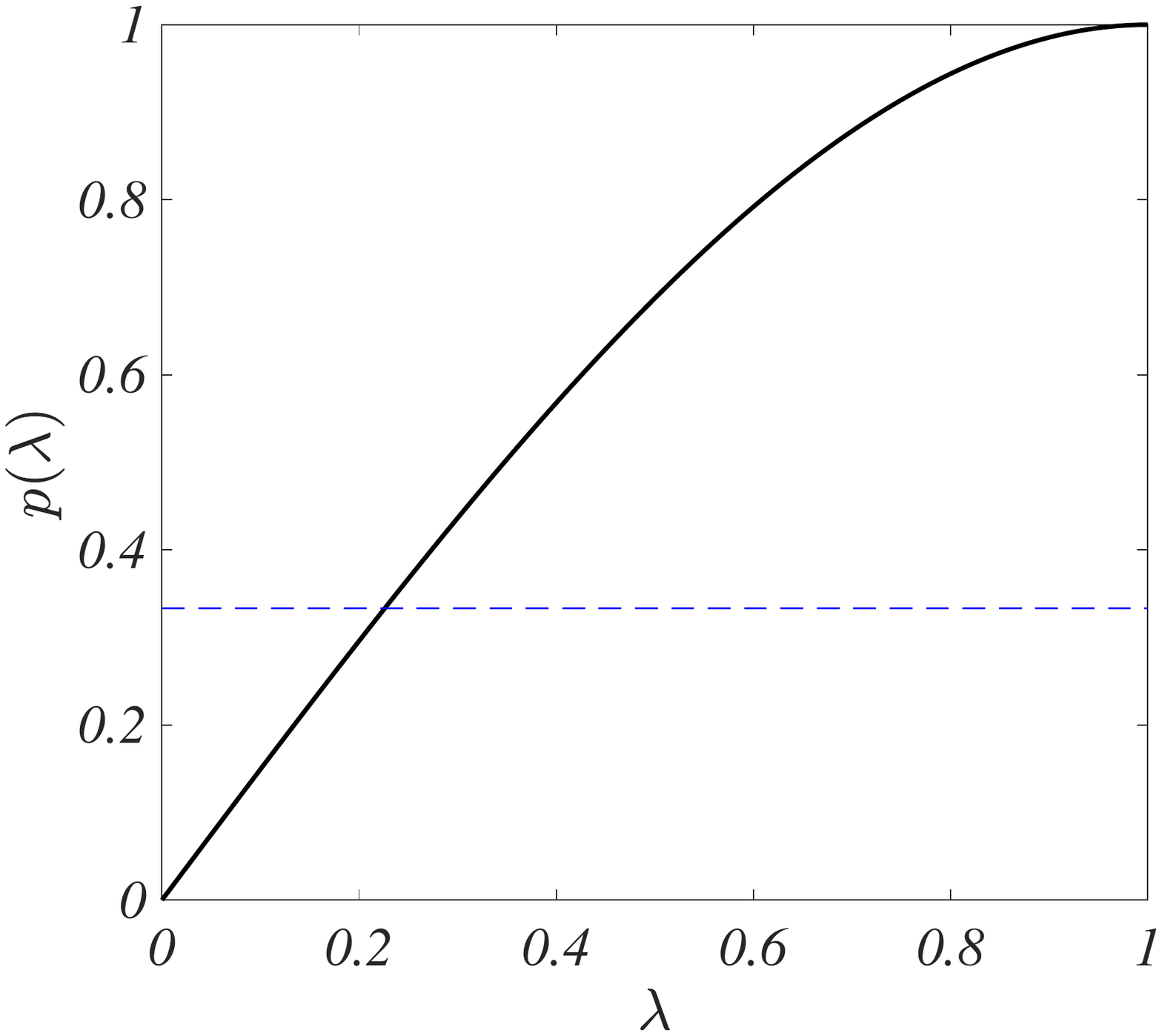}
\vspace{-1.5cm}
\caption{\textbf{Existence and uniqueness of the solution $p(\lambda)=f_n$}. We show the polynomial $p(\lambda)$ for $\lambda\in (0,1)$ (black curve) and a generic horizontal line set at $f_n\in (0,1)$ (blue dashed line).}
\label{fig:poly}
\end{figure}

In conclusion given $n$ we have to solve $n-1$ problems that we compactly write as
\begin{equation}
\label{eq:prob2solve}
p(\lambda)=\frac{3}{2}\lambda\left(1-\frac{\lambda^2}{3}\right)=f_n\, ,
\end{equation}
where $f_n\in\{1/n,\dots,(n-1)/n\}$. Before proceeding, let us observe that the above equation always admits one and only one solution in the interval $(0,1)$. Indeed, the polynomial $p(\lambda)$ vanishes at $\lambda=0$ and its first derivatives vanishes at $\lambda=\pm 1$. Further, it can be easily proven that it has a maximum at $\lambda=1$ where it reaches the value $p(1)=1$ and a minimum at $\lambda=-1$. The polynomial is thus steadily increasing from $0$ up to $1$. Hence, any constant horizontal line set at $f_n\in (0,1)$ intersects the polynomial in just one point belonging to the interval $(0,1)$ (see Fig.~\ref{fig:poly} for a graphical representation of this claim).
One can of course solve analytically the above equations, by making use of the explicit formulae for roots of the third order polynomials. Notice however that a good approximation for the first slices {can be obtained by setting} $\lambda_{app}=f_n \times \mathbf{2/3}$. {The adequacy of this latter ansatz is readily assessed by noticing that} $p(\lambda_{app})=f_n-f_n^3\times 4/27$ is very close to the required value, $f_n$, since the cubic term $f_n^3\times 4/27$ can be safely neglected.

Consider as an example $n=2$. We need to share every one fourth of watermelon into $2$ equal volume parts. According to the above recipe, we perform the cut at $\lambda_{app} a =a \times \mathbf{2/3} \times 1/2=a/3$, namely assuming a $1/3$-$2/3$ ratio of the semi-major axis. The above approximated criterion yields two slices with almost identical volumes, the relative difference in volume being as small as $4\%$.

For $n=3$ {we need to cut at $\lambda_1 a$ and $\lambda_2 a$, for $\lambda_1$ and $\lambda_2$ solutions of Eq.~\eqref{eq:root}. In analogy with the above, we can approximate $\lambda_1$ and $\lambda_2$ by using the $\mathbf{2/3}$-rule, yielding $\lambda^{(1)}_{app}=\mathbf{2/3}\times 1/3$ and $\lambda^{(2)}_{app}=\mathbf{2/3}\times 2/3$.} {Thus} the first slice can be obtained with a cut at $\lambda^{(1)}_{app} a =a\times \mathbf{2/3}\times 1/3 = 2a/9$ and its volume differs from the correct one, i.e. one third of the watermelon piece, by a small amount that we quantify in $1.6\%$. By using the same approximation to perform the second cut at $\lambda^{(2)}_{app} a =a\times \mathbf{2/3}\times 2/3 = 4a/9$, we obtain a slice $14\%$ {smaller (in volume) of the desired one.} The error is getting worse because the approximated cuts follow {a linear profile,} $\lambda_{app}=\mathbf{2/3}\times f_n$, while the {\em curvature} of the watermelon is responsible for the cubic term in $p(\lambda)$.

For a generic $n\geq 2$ one can compute the errors in volumes which follow the approximated cuts $\lambda^{(i)}_{app}a=a\times \mathbf{2/3}\times i/n$ for $i=1,\dots, n-2$. More precisely let us consider the ratio of the $i$-th approximated volume $V^{app}_{i}${, that is the volume obtained with a cut at $\lambda^{(i)}_{app}a$ (instead of $\lambda_{i}a$), computed from Eq.~\eqref{eq:root} evaluated at $\lambda^{(i)}_{app}a$,} with {$V$} the {initial} volume of the watermelon piece {(one fourth of the full volume)}:
\begin{equation*}
 \frac{V^{app}_{i}}{V}=p(\lambda^{(i)}_{app})-p(\lambda^{(i-1)}_{app})=\frac{i}{n}-\frac{4}{27}\frac{i^3}{n^3}-\frac{i-1}{n}+\frac{4}{27}\frac{(i-1)^3}{n^3}=\frac{1}{n}-\frac{4}{27}\frac{3i^2+3i+1}{n^3}\, .
\end{equation*}
One can immediately realise that the relative error grows with $i$ and {is} largest for the last cut, $i=n-1$.

\begin{figure}[ht]
\centering
\includegraphics[scale=0.4]{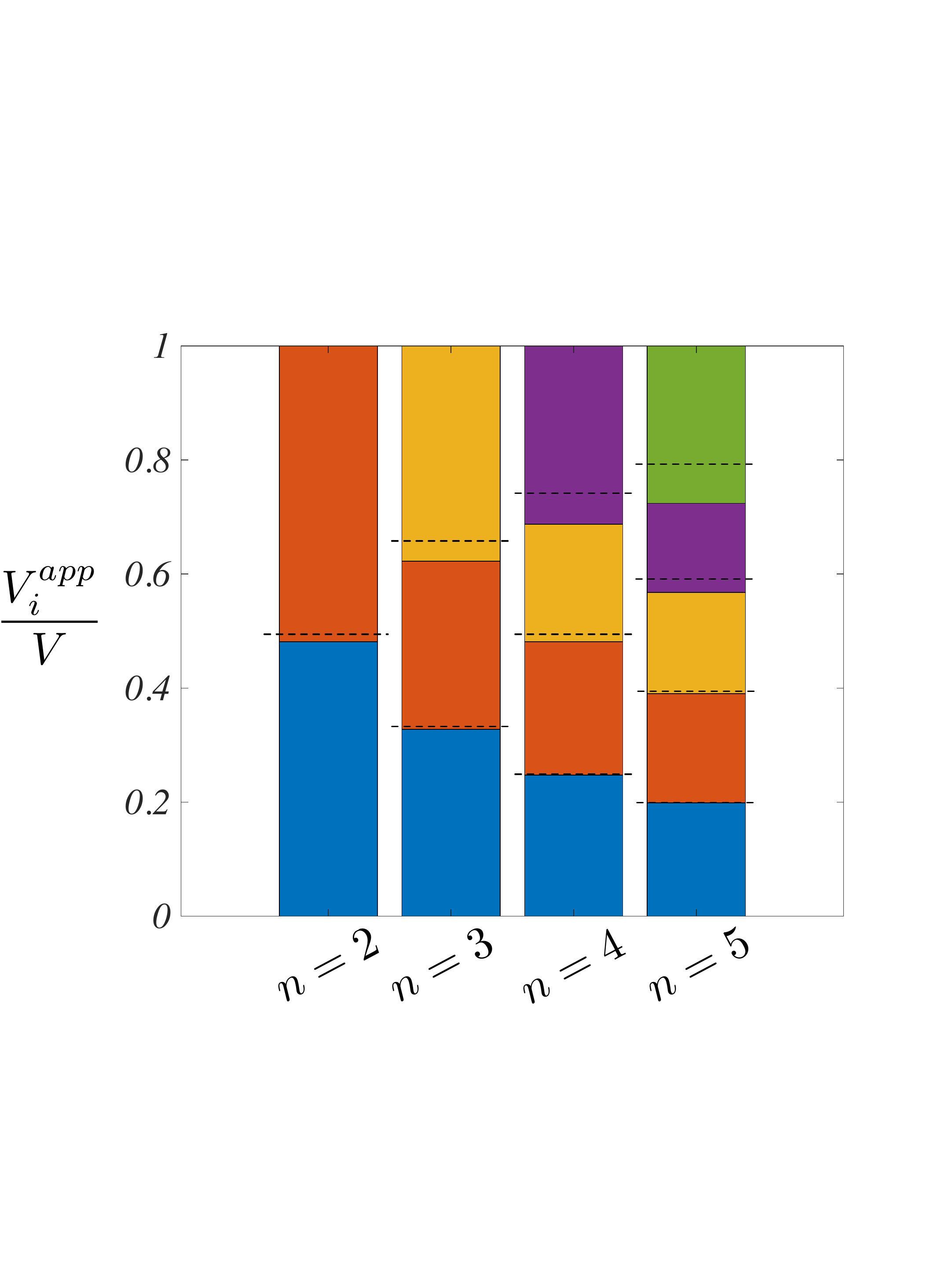}
\vspace{-1.5cm}
\caption{\textbf{Approximated volumes}. Each column represents the relative approximated volumes $V_i^{app}/V$ for a choice of $n$. The horizontal dashed lines represent the correct equal share for the corresponding $n$, i.e. $1/n$.}
\label{fig:volapp}
\end{figure}

The {suitability} of the proposed approximation can be evaluated upon inspection of Fig.~\ref{fig:volapp} where we report the approximated volumes for several values of $n$ as compared to their exact homologues (horizontal dashed lines). One can see that for all $n$ the first few approximated volumes are quite accurate, while the disagreement increases for further cuts at fixed $n$.


\section{The experiment}
\label{sec:exp}
To challenge our findings and the underlying hypotheses, (i.e. the watermelon can be correctly approximated by a spheroid and the approximated cuts provide equal volumes), we performed the experiment {and cut one fourth of a watermelon into}  $n$ {(supposedly equal volume)} portions. In line with the spirit of the note, we decided to use only tools available in our kitchen~\cite{Ficken1976,Bouquet2019} so as to make our experiment easily reproducible by everyone. The experiment has been filmed and a movie is available at~\cite{cocomeromovie}.

\begin{figure}[ht]
\centering
\includegraphics[scale=0.15]{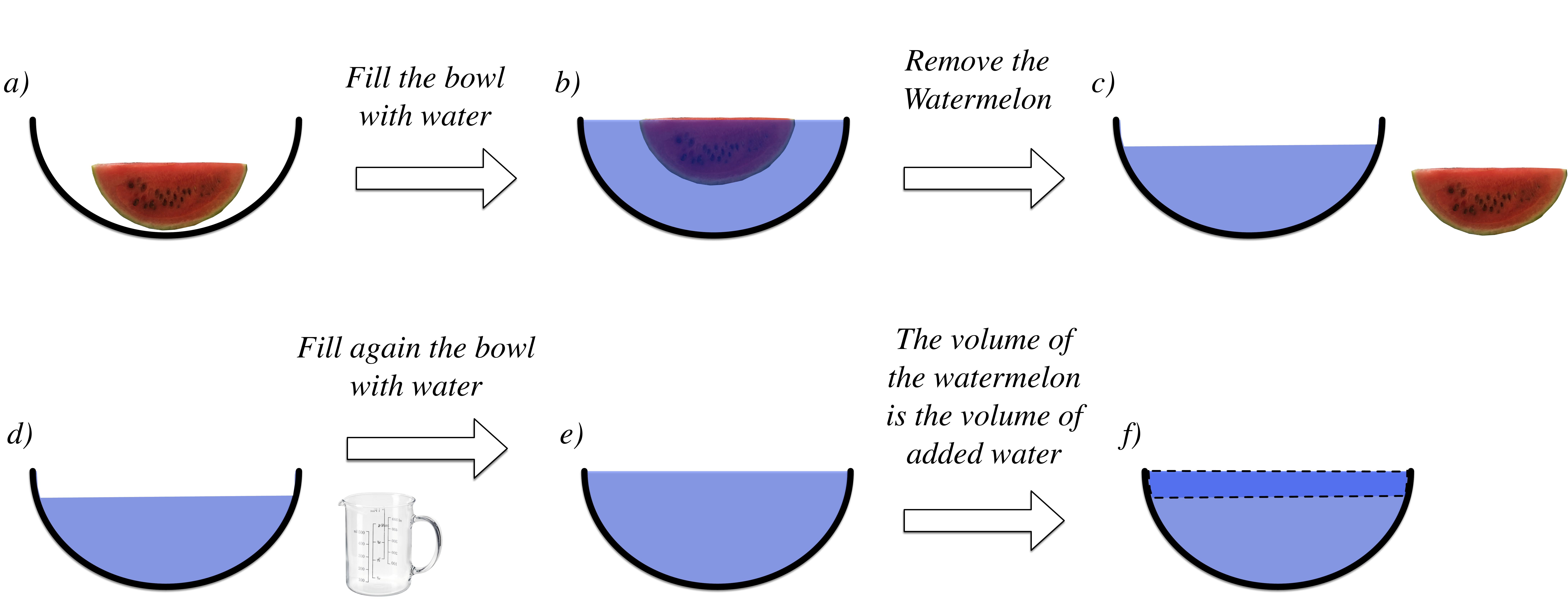}
\vspace{0.5cm}
\caption{\textbf{Measure a volume by using the Archimedes principles}. Panel a): insert in the bowl the watermelon whose volume is to be measured. Panel b): fill the bowl with water up to a reference mark. Panel c): remove the watermelon. Panels d)-e): to measure the drop in volume, fill again the bowl with water up to the reference mark. Panel f): the volume of the watermelon equals the added volume of water.}
\label{fig:ArchiPrinc}
\end{figure}

For the experiment we used a small watermelon weighting approximately $4\,\mathrm{Kg}$. Of course the results are independent from the watermelon size. The first step is to measure the watermelon volume. To accomplish this task we make use of the Archimedes principle~\cite{Archimedes,Thompson2008}: a body immersed in water, will displace a volume of water equal to the volume of the body. We cut the watermelon into two equal parts (longitudinal cut as shown in Fig.~\ref{fig:cocomero2}), we insert each part in a sufficiently large bowl and we fill it with water until a reference mark is reached. We then remove the watermelon: the volume of water in the bowl is hence lower (for the sake of completeness we refer the reader to the discussion at the end of this section). To measure this volume we use a measuring cup to fill again the bowl. Once the water level has reached again the initial reference mark, the added volume of water is equal to the volume of the watermelon (See Fig.~\ref{fig:ArchiPrinc}). In our case we got $\sim 1.9\, \mathrm{l}$ for each piece.

\begin{figure}[ht]
\centering
\includegraphics[scale=0.25]{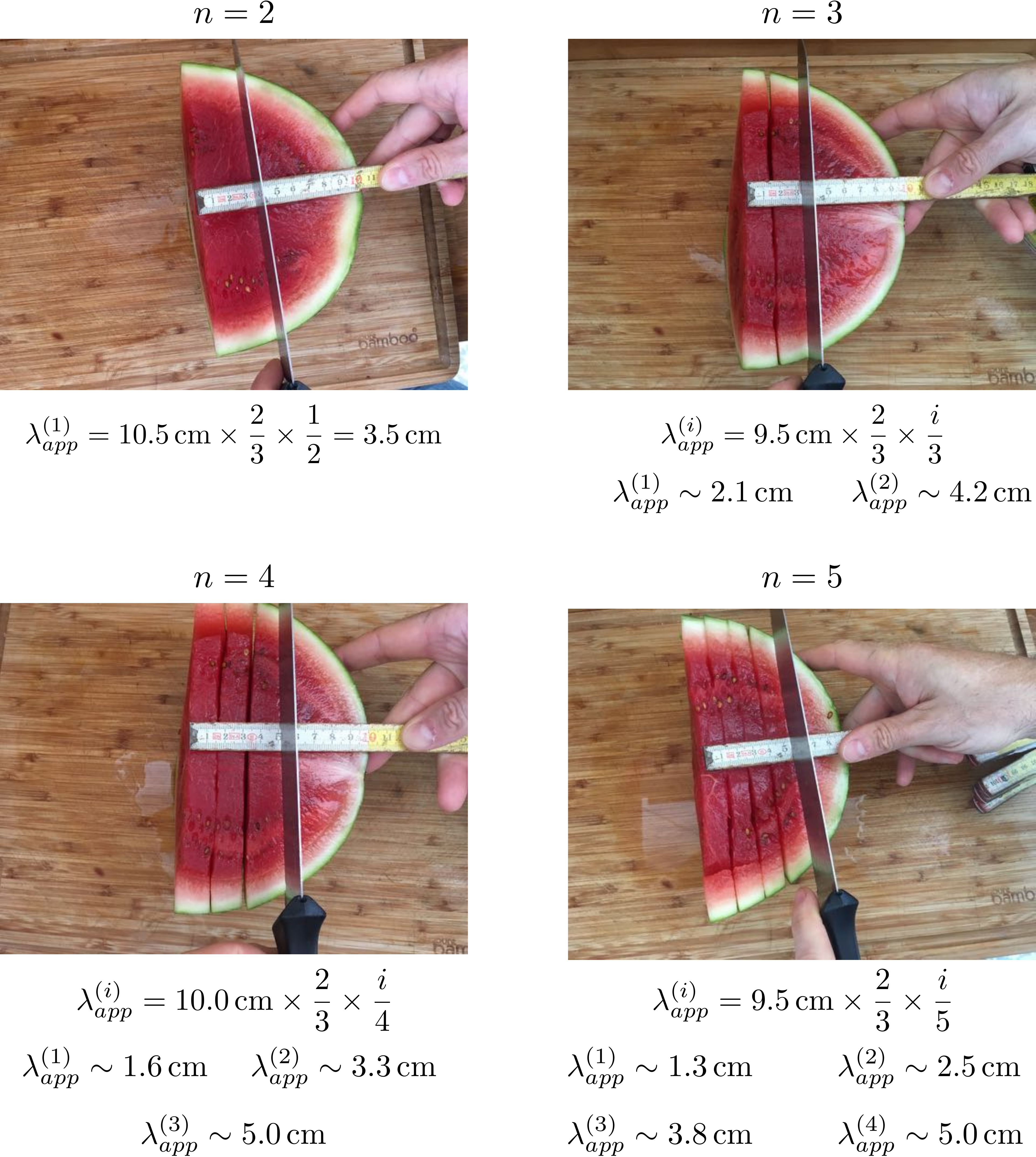}
\vspace{0.5cm}
\caption{\textbf{The approximated cuts realised in the experiment}. We report the approximated cuts realised in the experiment using the formula  $\lambda^{(i)}_{app}a=a\times \mathbf{2/3}\times i/n$, where $a$ is the semi-major axis of the spheroid (the watermelon) for several values of $n$: $n=2$ (top left panel), $n=3$ (top right panel), $n=4$ (bottom left panel) and $n=5$ (bottom right panel).}
\label{fig:ResCutExp}
\end{figure}

We are now ready to prepare the $n$ slices. To this end we cut the two half of watermelon into two parts each, yielding $4$ quarters. The first quarter is sliced into $n=2$ equal volume portions, from the second $n=3$, from the third $n=4$ and finally $n=5$ slices from the last one. In each case the positions of the cuts have been determined by using the approximation $\lambda^{(i)}_{app}a=a\times \mathbf{2/3}\times i/n$ for $i=1,\dots, n-2$ (see Fig.~\ref{fig:ResCutExp}).

The volumes of the cut slices have been measured by using the Archimedes principle; the results reported in the Table~\ref{tab:volexp} confirm the correctness of the $2/3$-rule in the computation of equal volume slices.
\begin{table}[th]
\begin{tabular}{c|c|c|c|c}   
  \hline
Slice number & $n=2$& $n=3$ & $n=4$  & $n=5$\\
  \hline
$1$ &  $490$& $300$ & $230$  & $170$\\
$2$ &  $510$& $290$ & $220$  & $170$\\
$3$ &$*$& $330$ & $230$  & $160$\\
$4$ &  $*$& $*$ & $240$  & $140$\\
$5$ &  $*$& $*$ & $*$  & $190$\\  
\end{tabular}
\caption{\textbf{Volumes of the slices}.
We report the volumes of the watermelon slices (expressed in $\mathrm{cm^3}$) obtained in the experiment. Each column corresponds to a choice of $n$, while the rows refer to the slice number. The symbol $*$ means that the slice is not allowed for the choice of $n$.}
\label{tab:volexp}
\end{table}

During the experiment we could {also} obtain an accessory result: a measure of the {\em density} of the watermelon, $\rho$, namely the quantity of matter contained in the (watermelon) volume. To do this we use a kitchen scale to weight the watermelon. This first half of watermelon weights $\sim 1.89\, \mathrm{Kg}$ for a volume of $\sim 1.95\, \mathrm{l}$; for the second half, $\sim 1.73\, \mathrm{Kg}$ and a volume of $\sim 1.81\, \mathrm{l}$. The density results thus (we take the average of the two measures)
\begin{equation*}
\rho\sim \frac{1}{2}\left(1.89 \, \mathrm{Kg}/ 1.95 \, \mathrm{l} +1.73 \, \mathrm{Kg}/ 1.81 \, \mathrm{l}\right)= 0.97\, \mathrm{Kg}/  \mathrm{l}=0.97 \, \mathrm{g}/  \mathrm{cm^3}\, .
\end{equation*}

Observe that the density is smaller than $1\,\mathrm{g}/  \mathrm{cm^3}$ which can be assumed to be the density of the water at the experimental conditions. Wa can conclude that the watermelon is less dense than water and thus it (almost) floats. Indeed the watermelon is composed for a large fraction by water and the weight of the few fibres (heavier than water) is compensated by the air (way lighter than water) trapped into the watermelon pulp. Let us observe that in~\cite{FoongLim2010} the density of the watermelon has been estimated to $0.94 \, \mathrm{g}/  \mathrm{cm^3}$, assuming a spherical watermelon and measuring its buoyancy.

Let us conclude this section with an observation about the application of the Archimedes principles. As we have seen the watermelon is slightly lighter than water and thus it can not be completely immersed; to measure the volume of the watermelon part we are interested in, we have thus to force it into the water.

\section{Conclusion}
\label{sec:conc}
We have provided a straightforward recipe to (transversally) cut a (spheroid) watermelon into an arbitrary number of pieces with equal volume. Interestingly enough, after the first few cuts dictated by the symmetry, i.e. divide by $2$ along the semi-major and minor axes, a ``new rule'' of the thumb emerges where the ratio $1/2$ is replaced by $\mathbf{2/3}\times 1/n$ where $n$ identifies the number of (equal volume) slices that one wants to recover.

Let us conclude that the same result, and thus an identical rule, holds true for a spherical watermelon. In the latter cases one could always invoke the spherical symmetry and thus cut slices with an angle of $2\pi /n$ (sort of apple wedges); however for large $n$ it can be difficult {correctly identify} such thin angles, while cutting vertically is normally easier.

The theory has been complemented with an experiment (a movie describing the experiment is also available~\cite{cocomeromovie}), realised by using tools available in our kitchen and thus hopefully reproducible in the students houses.

{As a final comment, notice that the relative distribution of (watermelon) skin has not been considered in the computation. The obtained equal volume slices display an uneven distribution of edible and non edible material. Accounting for this (i.e. producing equal volume slices of edible pulp) is in principle possible (as a straightforward generalisation of the method proposed) at the price of a more cumbersome computation which goes beyond the scope of this pedagogical note.}

\end{document}